\documentclass[conference]{IEEEtran}
\IEEEoverridecommandlockouts
\usepackage{cite}
\usepackage{amsmath,amssymb,amsfonts}
\usepackage{algorithmic}
\usepackage{booktabs}
\usepackage{multirow}
\usepackage{graphicx}
\usepackage{textcomp}
\usepackage{placeins}
\usepackage{xcolor}
\def\BibTeX{{\rm B\kern-.05em{\sc i\kern-.025em b}\kern-.08em
    T\kern-.1667em\lower.7ex\hbox{E}\kern-.125emX}}
\begin{document}

\title{A Category-Selective Encryption Framework for Privacy Protection of UAV Images
}

\author{
\IEEEauthorblockN{1\textsuperscript{st} Ting Yan}
\IEEEauthorblockA{
\textit{Dartmouth College}\\
Hanover, NH 03755, USA\\
ting.yan.gr@dartmouth.edu
}
\and
\IEEEauthorblockN{2\textsuperscript{nd} Minghui Shang}
\IEEEauthorblockA{
\textit{School of Information Science and Engineering}\\
\textit{University of Jinan}\\
Jinan 250022, China\\
2870062251@qq.com
}
\and
\IEEEauthorblockN{3\textsuperscript{rd} Haolang Sun}
\IEEEauthorblockA{
\textit{School of Information Science and Engineering}\\
\textit{University of Jinan}\\
Jinan 250022, China\\
lannistersun@126.com
}
\and
\IEEEauthorblockN{4\textsuperscript{th} Xu Wang}
\IEEEauthorblockA{
\textit{School of Information Science and Engineering}\\
\textit{University of Jinan}\\
Jinan 250022, China\\
ise\_wangx@ujn.edu.cn
}
\and
\IEEEauthorblockN{5\textsuperscript{th} Yamin Han*}
\IEEEauthorblockA{
\textit{Quancheng Laboratory}\\
Jinan 250100, China\\
hr-hanym@qcl.edu.cn
}
\and
\IEEEauthorblockN{6\textsuperscript{th} Mingze He}
\IEEEauthorblockA{
\textit{Department of Computer Science and Technology}\\
\textit{Sichuan Police College}\\
Luzhou, 646000, China\\
mingzebox@163.com
}
}

\maketitle

\begin{abstract}
Imagery captured by unmanned aerial vehicle (UAV) systems increasingly raises privacy concerns, as UAV aerial surveillance routinely captures sensitive objects such as pedestrians and vehicles alongside ordinary scene content. Existing privacy-preserving approaches typically employ class-agnostic saliency detection to indiscriminately obfuscate all visually salient regions, leading to unnecessary information loss and hindering downstream scene analysis. This paper proposes a category-selective encryption framework for privacy-preserving UAV surveillance. The core advance over prior class-agnostic approaches is semantic-level selectivity: the system first recognizes what is present in the scene via an 80-class detector, then encrypts only user-specified sensitive categories while leaving non-sensitive objects and background intact for downstream analytics. The framework supports two complementary encryption schemes: Thumbnail-Preserving Encryption (TPE), which applies adaptive pixelation with a learnable blend parameter and recovers content through a jointly trained Enhanced U-Net, and XOR stream cipher encryption, which offers parameter-free, mathematically exact reversibility. The architectural design, composite loss formulation, and joint training methodology are presented in detail, establishing a foundation for practical privacy-preserving UAV surveillance that balances protection strength and scene utility.
\end{abstract}


\section{Introduction}\label{sec:1}
With the rapid development of the low-altitude economy, UAV aerial photography, as a high-efficiency data acquisition technology with high-resolution imaging capability, has been widely used in agricultural monitoring \cite{daponte2019}, urban planning \cite{xu2023}, and environmental inspection \cite{gallacher2016}. However, such aerial images usually contain sensitive private information including human faces, pedestrians, and vehicle license plates. If these unprocessed images are leaked during air-to-ground wireless transmission or cloud storage, personal privacy will be exposed, which will lead to privacy breaches \cite{abu-khadrah2025}. Moreover, with the continuous advancement of computer vision technologies, deep learning algorithms can accurately identify human contours and behavioral activities in aerial images \cite{liu2019}.

To mitigate the privacy risks of aerial images, researchers have put forward a variety of image privacy protection schemes \cite{eltoukhy2025}. In recent years, a growing number of dedicated encryption algorithms have been developed specifically for UAV images. Shafique et al. \cite{shafique2023} proposed an encryption scheme combining DNA encoding and bit-plane extraction to secure high-resolution UAV-captured images during transmission. Zhou et al. \cite{zhou2024twodimensional} designed an encryption and compression algorithm based on two-dimensional hyperchaos for aerial images collected by agricultural UAVs. For smart agriculture scenarios, Zhou et al. \cite{zhou2024smart} constructed an image protection scheme that optimizes chaotic S-boxes via simulated annealing and affine transformation. Aiming at facial privacy protection in UAV remote sensing images, Wen et al. \cite{wen2024exploring} built a hybrid encryption framework integrating chaotic systems and DNA cryptography. Nevertheless, nearly all of the above schemes adopt a single fixed encryption mode. They fail to achieve discriminative selective encryption for semantic targets such as pedestrians, vehicles and public background areas, so they cannot adapt to diverse UAV patrol scenarios. Other mainstream privacy protection schemes rely on universal visual saliency detection, which applies blurring and encryption uniformly to all salient regions in an image \cite{Tu2020}. These schemes cause massive loss of valid scene information and severely interfere with subsequent data analysis. 

To meet the requirements of personal information protection laws and low-altitude data security regulations, there is an urgent need to develop a UAV aerial image privacy protection scheme that supports category-selective processing while balancing privacy protection and scene usability. To this end, this paper proposes a category-selective encryption framework for UAV images. The framework integrates multi-category object detection to achieve semantic-level selective encryption and incorporates two complementary and switchable encryption strategies. It constructs a complete closed-loop processing pipeline of  object detection, category-selective encryption, decryption and restoration, while maximizing the preservation of complete scene information under privacy security guarantees.

The main contributions of this paper are summarized as follows:

\begin{enumerate}
    \item The proposed framework employs a YOLO11 multi-category object detector to identify all 80 classes of targets in the image and applies encryption only to user-defined sensitive categories.
    \item The system integrates two pluggable and complementary encryption schemes, TPE and XOR stream cipher encryption. This design effectively accommodates the varying privacy protection requirements across different UAV aerial photography scenarios.
    \item A collaborative optimization mechanism covering detection, encryption, and restoration is established with a multi-dimensional composite loss function to significantly improve the decryption quality of the TPE scheme.

\end{enumerate}

The remainder of this paper is organized as follows. Section \ref{sec:2} reviews relevant studies. Section \ref{sec:3} elaborates on the specific implementation of the encryption framework we design. Section \ref{sec:sec4} presents experimental results and analysis. Finally, Section \ref{sec:sec5} concludes this paper and outlines future research directions.

\section{Related Works}\label{sec:2}
With the increasing popularity of UAV aerial photography applications, privacy security issues arising from aerial images have attracted widespread attention. This section introduces the TPE algorithm and XOR stream cipher encryption technology.

\subsection{TPE}
To reduce the risk of image data leakage in cloud environments, traditional schemes encrypt images into visually meaningless noise. However, this approach renders ciphertext images completely unreadable and fails to support cloud-side tasks such as image retrieval, management and analysis. To address this limitation, Wright et al. \cite{wright2015} proposed a novel encryption paradigm that balances privacy protection and image usability, namely TPE. A standard TPE scheme generally divides the original image into multiple sub-blocks and encrypts each sub-block independently while retaining the pixel average of each block to preserve the overall visual outline of the image. The trade-off between encryption security and image usability can be flexibly adjusted by modifying the block size. With the advancement of relevant research, existing TPE techniques are categorized into exact TPE \cite{peng2025,zhang2026} and approximate TPE \cite{zhang2022hftpe,an2025} according to whether the ciphertext thumbnail is strictly consistent with the original thumbnail. Early studies mostly adopted traditional signal processing and cryptographic operations, maintaining thumbnail structures via block permutation, pixel replacement and mean value preservation to strike a balance between security and usability. In recent years, deep learning has achieved remarkable progress in image modeling and generation, motivating researchers to adopt learnable models to fit the complex mapping between privacy protection effects and visual usability of images. Following this research line, Chai et al. \cite{chai2022} designed a GAN-based TPE-GAN scheme that greatly improves the usability of ciphertext images, yet there remains room for further optimization in the visual reconstruction quality of encrypted and decrypted images.

\subsection{XOR Stream Cipher Fundamentals}
XOR stream cipher implements encryption and decryption by performing bitwise XOR operations between the pseudo-random key stream and the binary sequences of image pixels. It relies on the inherent symmetric and reversible property of the bitwise XOR operation: for any pixel data \(d\) and key stream \(k\), the equation 
\begin{equation}
d \oplus k \oplus k = d
\end{equation}
always holds. The original image can be losslessly restored using the identical seed key \cite{mohammed2026}. With low computational overhead, this algorithm is well suited for lightweight terminal devices. However, traditional XOR stream ciphers only support full-image or fixed-region encryption and lack the capability of semantic target screening. After encryption, the overall visual information of the image is completely lost, making it difficult to balance privacy security and image usability.

\section{Proposed Scheme}\label{sec:3}
\begin{figure}
  \centering
  \includegraphics[width=\linewidth]{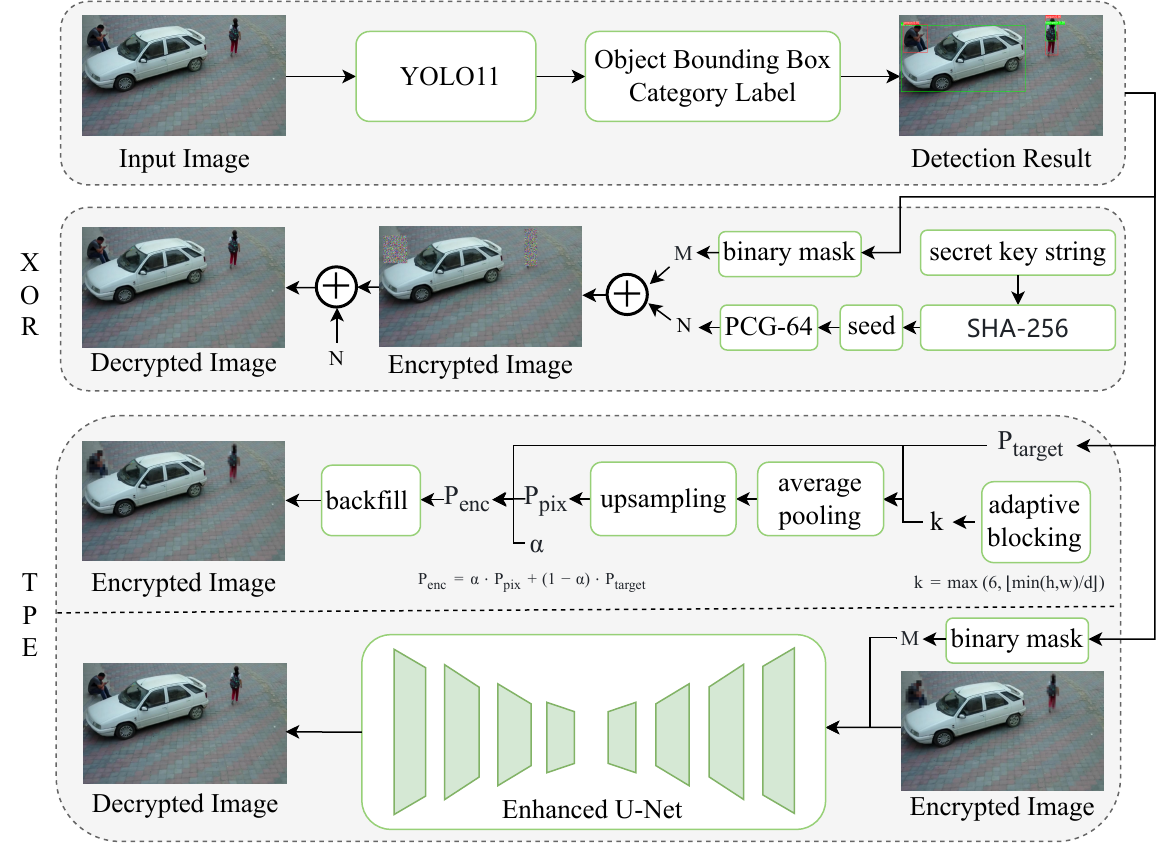}
  \caption{Overall Flowchart of the Proposed Scheme.}
  \label{fig:fig1}
\end{figure}

This section proposes a category-selective encryption framework oriented to object detection, which forms a closed-loop pipeline of object detection, category-selective encryption, decryption/image restoration as shown in Fig. \ref{fig:fig1}. The framework consists of three core modules: a multi-class object detection module based on YOLO11, a TPE module, and an XOR stream cipher encryption module. An independent decryption and restoration module is matched to the above encryption branches, and a multi-stage joint training strategy is specially designed only for the TPE branch. The detailed implementation schemes of each module are elaborated as follows.

\subsection{Object Detection Module}
This module adopts the lightweight multi-class detector YOLO11n to identify and categorize all objects within aerial scenes, which is fine-tuned on the VisDrone2019 dataset \cite{zhu2021}. The detector outputs bounding boxes for all detected targets. Users can customize a set of sensitive semantic categories to be encrypted at runtime; only objects falling into the designated sensitive classes are processed with encryption. Bounding boxes screened by the category filter are normalized to a unified coordinate range for subsequent mask generation, enabling semantic-aware selective encryption.
\subsection{Selective Encryption Schemes}
The proposed framework integrates two interchangeable encryption schemes with fundamentally different mechanisms, namely TPE and XOR stream cipher encryption. The two schemes share a unified input interface. Only sensitive pixel regions covered by detection boxes are modified, while all non-sensitive background pixels remain identical to the original image.
\subsubsection{TPE Encryption}
TPE performs pixelation transformation on sensitive target regions, and the original visual content can be restored by a well-trained Enhanced U-Net decoder. The encryption generates mosaic blur via average pooling and nearest-neighbor upsampling. A learnable blending coefficient \(\alpha\) is introduced to adjust the weight ratio between pixelated patches and original patches, which automatically balances privacy protection strength and image restoration quality during joint training. For each bounding box containing sensitive regions, the encryption pipeline consists of three steps:
\paragraph{Adaptive Block Size Calculation}
The kernel size of each block is dynamically calculated according to the width and height of the target box:
\begin{equation}
k = \max\left(k_{\min}, \left\lfloor\frac{\min(h, w)}{d}\right\rfloor\right)
\end{equation}
where the minimum block size \(k_{\min}=6\) and the scaling factor \(d=8\). Larger blocks are used for larger targets to achieve stronger mosaic blur, while distant small targets use finer blocks, ensuring consistent visual encryption effects for objects of all scales.
\paragraph{Pixelation Processing}
The image patch \(\mathbf{P}_{\text{target}}\) inside the bounding box is processed by average pooling with identical kernel size and stride k, then upsampled back to the original patch resolution \((h,w)\) via nearest-neighbor interpolation. Average pooling compresses each \(k\times k\) pixel block into a single average color value, and the subsequent upsampling recovers the spatial size of the region to form standard mosaic texture.
\paragraph{Learnable Blending Fusion}
The encrypted patch is obtained by linear weighted fusion of the pixelated patch and the original patch, where the weight is controlled by the learnable parameter \(\alpha\):
\begin{equation}
\mathbf{P}_{\text{enc}} = \alpha \cdot \mathbf{P}_{\text{pix}} + (1 - \alpha) \cdot \mathbf{P}_{\text{target}}
\end{equation}
\(\alpha\) is constrained within the range \([0,1]\) in the forward propagation. When \(\alpha=1\), the region is fully pixelated with the strongest privacy protection; when \(\alpha=0\), the patch is identical to the original image without any encryption effect.
\subsubsection{XOR Stream Cipher Encryption}
The XOR stream cipher realizes encryption relying on the involution property of the bitwise XOR operation. Deterministic pseudo-random noise is generated from the user-specified secret key to obfuscate pixels in sensitive regions, which achieves mathematically lossless reversibility without any neural network for restoration. The user-defined key string is mapped to a fixed-length seed through SHA-256 hashing. The seed initializes the PCG-64 pseudo-random number generator to produce a three-channel uniform noise map with the same resolution as the input image. The whole noise map is generated at one time rather than box by box, which guarantees consistent encryption results even when target boxes overlap. A binary mask \(\mathbf{M} \in \{0, 255\}^{H \times W}\) is constructed, where pixels inside sensitive boxes are set to 255 and background pixels are set to 0. Bitwise XOR operation is only conducted on pixels marked by the mask:
\begin{equation}
\mathbf{Y}[i, \mathbf{M} = 255] = \mathbf{X}[i, \mathbf{M} = 255] \oplus \mathbf{N}[\mathbf{M} = 255]
\end{equation}
where \(\mathbf{X}\) and \(\mathbf{Y}\) denote the original uint8-format image and encrypted image respectively. Background pixels uncovered by the mask retain their original values.

\subsection{Decryption and Restoration Module}
This paper designs supporting decryption pipelines for TPE and XOR encryption branches to restore original content from encrypted images.
\subsubsection{TPE Decryption}
The learnable mosaic blur of the TPE module lacks a straightforward mathematical inverse transformation. An enhanced U-Net decoder is thus utilized to recover texture information from masked sensitive regions. The network receives the encrypted image \(\mathbf{I}_{\text{enc}}\) as input. Its multi-scale skip connections suppress mosaic artifacts and restore target details without altering background pixels, yielding the restored output \(\mathbf{I}_{\text{rest}}\). Joint training is performed on the U-Net decoder and TPE module to maintain reliable restoration quality for targets of diverse scales and different encryption intensity \(\alpha\).
\subsubsection{XOR Stream Cipher Decryption}
Relying on the involution property of bitwise XOR, decryption follows the same workflow as encryption. The original secret key regenerates PCG-64 pseudo-random noise, and the encryption mask serves to localize sensitive pixels. Performing XOR once more over masked regions recovers lossless uint8 pixel values without neural network computation, which enables lightweight decryption.
\subsubsection{Post-processing Coordinate Inversion}
After image restoration of two decryption branches, metadata cached in LetterBox preprocessing (scaling ratio, padding offsets and original resolution) is loaded to implement geometric inverse transformation. Symmetric padding is removed and images are rescaled to original size, so as to eliminate image deformation caused by preprocessing.

\subsection{Joint Training Strategy}
A modular joint training scheme is adopted for the detection, encryption and decryption modules. Gradient backpropagation optimization is only performed on the TPE encryption branch to balance detection accuracy, privacy protection performance and image restoration quality.
\subsubsection{Parameter Freezing and Update Rules}
The weights of the pre-trained YOLO11n detection network are fully frozen throughout joint training, so that the target localization and classification performance will not be disturbed by the encryption and restoration task. Only two types of parameters are updated during training: the learnable blending coefficient \(\alpha\) of TPE and all weights of the Enhanced U-Net restoration network. This strategy drastically reduces the number of parameters to be optimized and cuts down GPU memory consumption during training. The binary mask \(\mathbf{M}\) generated in the encryption stage is reused for loss calculation. Only sensitive pixels covered by the mask participate in loss computation and gradient backpropagation, while background pixels are excluded from optimization to guarantee consistency between the background area and the original image.
\subsubsection{Loss Function Design}
The overall joint loss is composed of a weighted sum of reconstruction loss and encryption regularization loss:
\begin{equation}
\mathcal{L}_{\text{total}} = \mathcal{L}_{\text{rec}} + \lambda_{\alpha}\mathcal{L}_{\alpha}
\end{equation}
where \(\lambda_{\alpha}\) denotes the regularization weight, which is set to 0.6 in the experiments.
\paragraph{Mask-Constrained Reconstruction Loss \(\mathcal{L}_{\text{rec}}\)}
L1 loss is utilized to optimize the restoration quality of sensitive regions, and pixel-wise error is only calculated within the masked encryption area:
\begin{equation}
\mathcal{L}_{\text{rec}} = \big\|(\mathbf{I}_{\text{rest}}-\mathbf{I}_{\text{gt}})\odot\mathbf{M}\big\|_1
\end{equation}
\(\mathbf{I}_{\text{gt}}\) represents the original clear ground-truth image, and \(\mathbf{I}_{\text{rest}}\) stands for the restored image output by the network.
\paragraph{Encryption Strength Regularization Loss \(\mathcal{L}_{\alpha}\)}
This loss constrains the blending coefficient to approach 1, preventing the model from excessively weakening mosaic encryption merely to minimize reconstruction loss and maintaining effective privacy shielding capability:
\begin{equation}
\mathcal{L}_{\alpha} = 1-\alpha
\end{equation}

\section{Experiment Analysis}\label{sec:sec4}
All experiments are carried out on the public UAV aerial dataset VisDrone \cite{zhu2021}. The YOLO11n detection backbone loads weights fine-tuned on VisDrone and remains fixed throughout the joint training process. For the TPE module, the minimum block size \(k_{\min}=6\) and scaling factor \(d=8\) are adopted. The enhanced U-Net incorporates SE channel attention with a channel compression ratio \(r=8\). The AdamW optimizer is utilized for training with an initial learning rate of \(1\times10^{-4}\) and a batch size of 8. The regularization weight of the loss function is set to \(\lambda_{\alpha}=0.6\). The model is trained for a total of 120 epochs with an early stopping strategy enabled.

\subsection{Visual Effect Analysis}
\begin{figure}
  \centering
  \includegraphics[width=\linewidth]{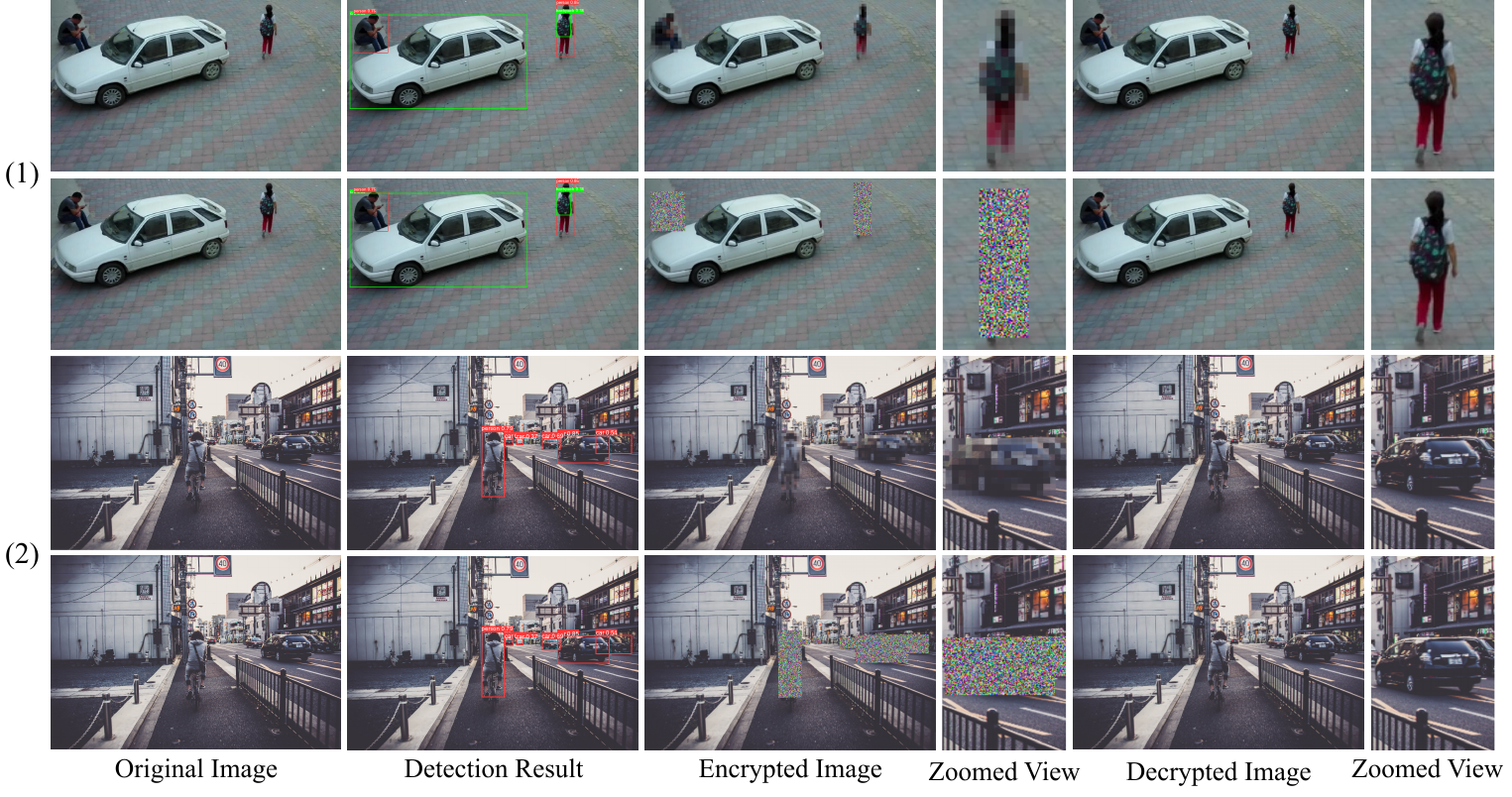}
  \caption{Visual comparison of two encryption strategies for UAV aerial images. Row (1) encrypts only pedestrians and Row (2) encrypts pedestrians together with vehicles. In each group, the top row corresponds to TPE encryption, and the bottom row corresponds to XOR encryption.}
  \label{fig:fig2}
\end{figure}

Two encryption algorithms are applied to two sensitive target filtering strategies in this section: encrypting pedestrians only, and encrypting both pedestrians and vehicles simultaneously. Visual results presented in Fig. \ref{fig:fig2} reveal that both algorithms exclusively encrypt sensitive regions while leaving background areas unchanged. For the TPE scheme, sensitive targets are fully occluded by adaptive mosaic blur. This mechanism makes private information such as human faces and vehicle appearances unrecognizable without introducing any distortion to the background. After decryption with the enhanced U-Net network, the textures and contours of encrypted targets can be satisfactorily reconstructed. In contrast, the XOR encryption branch covers sensitive regions with pseudo-random noise to achieve favorable visual obfuscation. Provided the correct secret key is available, strict lossless pixel-level image reconstruction can be accomplished. Moreover, both encryption schemes avoid blur, color shift and other background artifacts, and remain effective in challenging aerial scenes with sensitive targets visually close to background tones.
\begin{figure*}
  \centering
  \includegraphics[width=\linewidth]{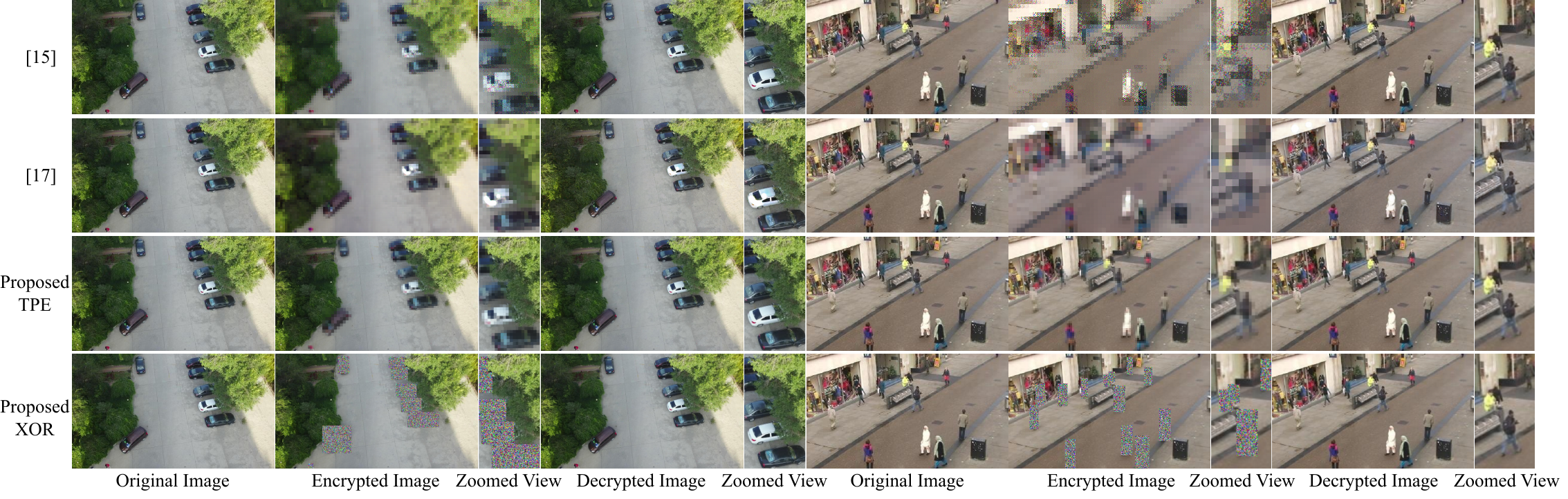}
  \caption{Visual Comparison of Visual Effects Between the Proposed scheme and TPE schemes}
  \label{fig:fig3}
\end{figure*}
\subsection{Image Quality Analysis}
To quantitatively evaluate the reconstruction performance of the two encryption schemes, we compute the Peak Signal-to-Noise Ratio (PSNR) and Structural Similarity Index Measure (SSIM) between the original aerial images and decrypted images on the VisDrone test set to compare the image restoration performance of TPE and XOR encryption. Higher PSNR and SSIM values indicate less distortion in decrypted images and better visual consistency with the original images. As can be seen in Table \ref{tab:ssim_psnr}, XOR encryption achieves an SSIM of 1 and infinite PSNR under both encrypted region settings, which demonstrates that XOR enables lossless pixel-level reconstruction without any image distortion after decryption. In contrast, TPE encryption introduces slight reconstruction degradation. The decryption quality remains favorable when only pedestrians are encrypted, while the restoration performance degrades when both pedestrians and vehicles are encrypted due to the expanded encrypted area.
\begin{table}[htbp]
\centering
\caption{PSNR and SSIM Quantitative Metrics of TPE and XOR under Different Encrypted Regions}
\label{tab:ssim_psnr}
\begin{tabular}{lccc}
\toprule
& Encrypted Region& SSIM& PSNR\\
\midrule
\multirow{2}{*}{TPE}& person& 0.9792& 38.48dB\\
& person \& car& 0.9734& 35.26dB\\
\multirow{2}{*}{XOR}& person& 1& inf\\
& person \& car& 1& inf\\
\bottomrule
\end{tabular}
\end{table}

\subsection{Comparative Analysis}
Existing saliency-driven privacy schemes, traditional region-selective encryption and basic object anonymization share two inherent limitations. Category-customized encryption for designated semantic targets cannot be realized, and most alternatives only support irreversible visual masking without complete high-fidelity recovery pipelines.

Two classic full-image TPE algorithms, HF-TPE \cite{zhang2022hftpe} and TPE-GAN \cite{chai2022}, are taken as baselines for fair performance benchmarking of the proposed TPE and XOR encryption modules. As illustrated in Fig. \ref{fig:fig3}, both baselines implement full-image encryption that triggers visual distortion on non-sensitive regions such as roads and buildings. Such distortion destroys inherent aerial scene features, impairs the performance of subsequent scene recognition tasks, and introduces substantial computational overhead. A favorable balance between privacy preservation and image utility is attained by the presented framework through encryption limited to predefined sensitive objects only. Reconstruction quality of all algorithms is quantified in Table \ref{tab:ssim_psnr2} with SSIM and PSNR metrics, where all approaches are categorized into cryptographic and deep learning-based groups. Satisfactory reconstruction performance is obtained by HF-TPE among cryptographic schemes. Benefiting from invertible XOR operations, the propsed XOR scheme branch delivers perfect lossless pixel reconstruction with an SSIM value of 1 and infinite PSNR. For deep learning-based TPE-GAN, its GAN encoder-decoder architecture generates severe reconstruction artifacts and deteriorates the quality of decrypted outputs. By contrast, the proposed TPE scheme achieves outstanding reconstruction fidelity and effectively alleviates such severe distortion. Overall, redundant encryption over background regions is eliminated by both presented schemes. Furthermore, superior reconstruction performance against corresponding baselines within each technical category is achieved by the two proposed encryption branches.

\begin{table}[htbp]
\centering
\caption{PSNR and SSIM Quantitative Metrics of Different Encryption Schemes under Various Encrypted Regions}
\label{tab:ssim_psnr2}
\begin{tabular}{lccc}
\toprule
 & Encrypted Scheme& SSIM& PSNR\\
\midrule
 {HF-TPE}{\cite{zhang2022hftpe}}& cryptography-based& 0.9996& 55.40dB\\
 {Proposed-XOR}& cryptography-based&  1& inf\\
 {TPE-GAN}{\cite{chai2022}}& deep learning-based& 0.8699& 28.07dB\\
 {Proposed-TPE}& deep learning-based& 0.9734& 35.26dB\\ \bottomrule
\end{tabular}
\end{table}
\subsection{Detection Performance and Privacy Implications}
Quantitative reliability tests on the VisDrone test set \cite{zhu2021} show that the pre-trained model achieves mAP@0.5 = 27.6\% and mAP@0.5:0.95 = 15.5\%, which is standard for lightweight detectors in complex UAV scenes with dense small objects and varying altitudes. Although distant tiny targets have limited recognizable details due to low resolution, privacy leakage may still occur through cross-frame contextual reasoning and temporal correlation analysis. Future research  will mitigate these risks through more robust detection pipelines or unified joint optimization of detection and encryption.  

\subsection{Efficiency Analysis}
Network parameter counts and average processing latencies of the proposed XOR and TPE modules are reported in Table \ref{tab:time}. The XOR pipeline is a handcrafted lightweight bitwise operation without neural network components, achieving only 0.5 ms latency for both encryption and decryption under the unified experimental environment. In the TPE branch, encryption and decryption show a clear latency gap: encryption uses only a manually defined hyperparameter \(\alpha\) to control adaptive blurring intensity in sensitive areas, resulting in a low latency of approximately 1 ms. In contrast, TPE decryption employs a deep convolutional reconstruction network with 50.29 million trainable parameters to recover distorted texture details, incurring substantial matrix computation and memory access overhead and a latency of around 30 ms. Benefiting from target-selective encryption, both modules restrict computation to YOLO11n sensitive bounding boxes, thereby reducing the number of processed pixels.
\begin{table}[htbp]
\centering
\caption{Network Parameter Count and Average Processing Latency of the Proposed XOR and TPE Modules}
\label{tab:time}
\begin{tabular}{lccc}
\toprule
& Stage& Network Parameter Count& Average Latency \\
\midrule
\multirow{2}{*}{TPE}& Encryption& — (hyperparameter \(\alpha\))& $\sim$1ms
\\
& Decryption& 50.29M& $\sim$30ms\\
\multirow{2}{*}{XOR}& Encryption& N/A& $\sim$0.5ms
\\
& Decryption& 
N/A& $\sim$0.5ms
\\ \bottomrule
\end{tabular}
\end{table}

\FloatBarrier
\section{Conclusion}\label{sec:sec5}
This paper designs a dual-mode category-selective image encryption framework integrated with object detection. YOLO is utilized to locate sensitive targets, and TPE adaptive pixelation encryption and XOR stream cipher encryption are separately adopted to realize regional privacy protection.
The TPE scheme leverages a learnable blending coefficient to balance the masking effect and image readability, and relies on an enhanced U-Net for texture restoration; the XOR scheme achieves lightweight lossless encryption and decryption. Experimental results demonstrate that the proposed scheme only encrypts targets of designated categories, which drastically reduces redundant computation. It simultaneously delivers visual privacy protection while preserving basic image content, making it suitable for real-time image transmission scenarios of UAVs. Future work will further optimize the network inference speed and encryption security strength, expand the applicable scope of the algorithm, and adopt more robust detection and joint detection-encryption optimization strategies to mitigate privacy leakage risks caused by undetected small, occluded or poorly illuminated sensitive objects.

\section*{Acknowledgment}
This work was supported in part by the Sichuan Science and Technology Program under Grant No. 2024NSFSC2045, and in part by the Research Project of Quancheng Laboratory, China under Grant No. QCL20250303.

\bibliographystyle{IEEEtran}
\bibliography{references}

\end{document}